\documentclass[sigconf]{acmart}

\usepackage{amsmath}
\usepackage{subcaption}

\AtBeginDocument{%
  }

\copyrightyear{2025}
\acmYear{2025}
\setcopyright{cc}
\setcctype{by}
\acmConference[WWW Companion '25]{Companion Proceedings of the ACM Web Conference 2025}{April 28-May 2, 2025}{Sydney, NSW, Australia}
\acmBooktitle{Companion Proceedings of the ACM Web Conference 2025 (WWW Companion '25), April 28-May 2, 2025, Sydney, NSW, Australia}
\acmDOI{10.1145/3701716.3715517}
\acmISBN{979-8-4007-1331-6/25/04}

\newcommand\method{\texttt{FNDCD}}

\begin{document}

\title{Unseen Fake News Detection Through Casual Debiasing}

\author{Shuzhi Gong}
\orcid{0009-0007-9289-9015}
\affiliation{%
\institution{The University of Melbourne}
\city{Melbourne}
\state{VIC}
\country{Australia}}
\email{shuzhig@student.unimelb.edu.au}

\author{Richard Sinnott}
\orcid{0000-0001-5998-222X}
\affiliation{%
\institution{The University of Melbourne}
\city{Melbourne}
\state{VIC}
\country{Australia}}
\email{rsinnott@unimelb.edu.au}

\author{Jianzhong Qi}
\orcid{0000-0001-6501-9050}
\affiliation{%
  \institution{The University of Melbourne}
  \city{Melbourne}
  \state{VIC}
  \country{Australia}
}
\email{jianzhong.qi@unimelb.edu.au}

\author{Cecile Paris}
\orcid{0000-0003-3816-0176}
\affiliation{%
  \institution{Data61, CSIRO}
  \city{Sydney}
  \state{NSW}
  \country{Australia}}
\email{Cecile.Paris@data61.csiro.au}

\renewcommand{\shortauthors}{Shuzhi Gong, Richard Sinnott, Jianzhong Qi, \& Cecile Paris}

\begin{abstract}
The widespread dissemination of fake news on social media poses significant risks, necessitating timely and accurate detection. However, existing methods struggle with unseen news due to their reliance on training data from past events and domains, leaving the challenge of detecting novel fake news largely unresolved.

To address this, we identify biases in training data tied to specific domains and propose a debiasing solution \method. Originating from causal analysis, \method~employs a reweighting strategy based on classification confidence and propagation structure regularization to reduce the influence of domain-specific biases, enhancing the detection of unseen fake news. Experiments on real-world datasets with non-overlapping news domains demonstrate \method's effectiveness in improving generalization across domains.

\end{abstract}

\ccsdesc[500]{Security and privacy~Social aspects of security and privacy}
\ccsdesc[500]{Computing methodologies~Artificial intelligence}

\keywords{Fake News Detection, Graph Neural Network, Generalization}


\maketitle

\section{Introduction}
The proliferation of social media has accelerated the spread of both accurate and misleading information. Early and reliable detection of fake news is thus crucial to minimizing its harmful societal impact. Recent advances in fake news detection have utilized graph-based techniques, especially Graph Neural Networks (GNNs), to model news propagation patterns and extract critical insights that might be used for identifying misinformation~\cite{gong2023survey}. However, these approaches typically assume that both training and testing data share the same underlying distribution (the  ``i.i.d.'' assumption). The trained models  often embed biases and noise present in the training data, leading to mis-classifications during model inference~\cite{liu2021towards-ood-generalisation-survey}. This means that the model has not been trained on representative data of the fake news to be detected. In reality, fake news has often never been seen before and originates from new domains, which poses significant challenges in generalizing models trained on known distributions (so called \emph{in-distribution}) to novel and unseen distributions (so called \emph{out-of-distribution}, OOD). 

Some studies have tried to identify content-independent propagation patterns to detect fake news across different news domains~\cite{ma2018rumor,bian2020rumor,gong2023fake}, however, more recent work suggests that both content and propagation structures may differ significantly between news domains~\cite{min2022divide}. When considering the training  and testing data from a source domain and target domain, domain adaptation approaches are often used ~\cite{li-acmmm-2023improving,lin-naacl-2022-aclr}. They attempt to address this issue by fine-tuning models using a small amount of labeled target domain data. However, such labeled data is not always available in real-world scenarios where the fake news has not been seen before, e.g. the breaking COVID-19 event. Furthermore, target domains can evolve quickly and there may not be enough time to label such new data before fake news spreads. Real fake news detection can thus be regarded as an out-of-distribution generalization task. 


To address these challenges, we propose \textbf{F}ake \textbf{N}ews \textbf{D}etection by 
\textbf{C}ausal \textbf{D}ebiasing (\method), a novel approach designed for zero-shot unseen domain fake news detection. Through causal analysis (see next section), we identify that the existence of biased training samples restricts the trained models' generalization performance on unseen data from new domains. A self-supervised weighting strategy is designed according to the news content, news propagation pattern, and existing labels. Through re-weighting, the contribution from the biased data during model training can be reduced to improve the trained model's cross-domain generalization and overall ability to tackle nascent fake news challenges.

Extensive experiments are conducted with four datasets used for unseen fake news detection. We show that \method~outperforms state-of-the-art models. \method~also provides interpretability and insights about the data and its potential for bias, allowing improvements for future news data collection and processing.

\section{Problem Formulation}\label{sec:problem}
\textbf{Unseen fake news detection.}
We consider graph-based fake news detection using propagation graphs (trees), comprising source news posts, comments and reposts.
Given a training dataset $D^{tr} = \{\mathcal{G}_{k}^{tr}, y_{k}^{tr}\}_{k=1}^{\mathcal{N}_{tr}}$, where $\mathcal{G}_{k}^{tr}$ is the $k$-{th} training propagation graph, $y_{k}^{tr}$ is its label, and $\mathcal{N}^{tr}$ is the number of training samples, the aim is to train a model using  $D^{tr}$ for optimal performance on unseen testing data $D^{te}=\{\mathcal{G}^{te}_{k}, y^{te}_{k}\}_{k=1}^{\mathcal{N}^{te}}$. Distribution shifts typically exist between $D^{tr}$ and $D^{te}$ when they are collected from different news domains at different times. In the unseen fake news detection setting, the features and labels of $D^{te}$  are unavailable during training.

For any data sample $\mathcal{G}=\langle \mathbf{X}, \mathbf{A} \rangle$ from $D^{tr}$ or $D^{te}$, the propagation graph $\mathcal{G}$ is composed of news posts, comments and reposts represented by node features $\mathbf{X}$, and user interactions (comments and reposts) represented by graph edges $\mathbf{A}$. To convert the raw news text into graph features, we use a pre-trained RoBERTa model ~\cite{liu2019roberta}. 

\textbf{Distribution shifts in fake news detection.} 

It has been established that distribution shifts often exist between  training data $D^{tr}$ and testing data  $D^{te}$~\cite{min2022divide,ran-aaai-2023-unsupervised}. In the context of misinformation and its propagation, these shifts can be characterised from three perspectives: shifts in textual content $\mathbf{X}$, shifts in propagation structure $\mathbf{A}$, and shifts in the label-feature correlations $p(y|\mathbf{X}, \mathbf{A})$. The shifts in textual content occur when the news and associated comments pertain to different news topics potentially across different news domains. For example, political news often involves vocabulary related to countries and politicians, whereas COVID-19 posts focus more on medical information. The shifts in propagation structure reflect the variation in propagation graphs between different news domains. Shifts in label-feature correlations arise when similar embeddings from different domains that might be extracted by traditional graph encoders, give rise to contrasting labels. This correlation shift presents a significant challenge for fake news detection for unseen news domains.

To address these shifts, a causal analysis is considered.

\begin{figure}[tbh]
  \centering
  \includegraphics[width=0.6\linewidth]{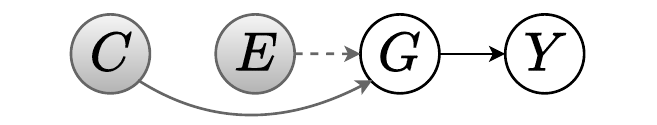}
  \caption{Structure of the causal model used for training cross-domain fake news detection. $C$: Causal information that supports the correct classification; $E$: Spurious environment-biased information harming the classification; $G$:  Observed graph features; $Y$: Associated veracity label. The grey and white variables represent the degree of observability (unobserved is grey and observed is white).}
  \Description{The SCM.}
  \label{Fig.SCM_of_training}
\end{figure}

\textbf{Causal analysis.}
\label{Sec.causal_analysis}
We hypothesize that generalization to unseen news domains is hindered by biases in the training set, such as biases toward specific events like the US presidential election (\emph{environment-bias}). To model this, we use a structural causal model (SCM)~\cite{pearl2009causality}, shown in Fig.\ref{Fig.SCM_of_training}, where nodes represent variables and edges represent causal effects. The observed propagation graph $G$ is generated from two latent variables: the causal variable $C$ and the environment-biased variable $E$. The label $y$ is predicted based on $G$, influenced by both $C$ and $E$.

Training on environment-biased samples embeds spurious correlations ($E \rightarrow G \rightarrow Y$), leading to sub-optimal performance on OOD data. These spurious correlations interfere with the causal relationship ($C \rightarrow G \rightarrow Y$). To address this, environment-biased samples must be identified and down-weighted, ensuring that only causal effects are preserved for accurate OOD classification.

\textbf{Rescue of probability.} 
\label{Section.rescue_of_probability}
Inspired by a recent work~\cite{li2022graphde}, the news data generation can be described by the joint probability of several variables: the textual content $\mathbf{X}$; the structure $\mathbf{A}$ of the propagation graph $\mathcal{G}$, the veracity label $\mathbf{y}$ and the environment variable $\mathbf{e}$.  


Here, the environment variable $\mathbf{e}$ is treated as independent because the other variables originate from it (i.e., the variety of domains/topics causes the distribution shifts). The news content $\mathbf{X}$ should depend on $\mathbf{e}$, yet, for simplicity, we use a domain-adaptive pre-trained language model (DA-PLM) to extract the news content features such that $\mathbf{X}$ is disentangled from $\mathbf{e}$. Considering the homophily principle theory~\cite{muller2012neural} linking the probability depending on some inherent similarity between nodes, we assume that users sharing similar interests are more likely to interact.  Variable $\mathbf{A}$ is hence defined as depending on $\mathbf{X}$ and $\mathbf{e}$. Finally, the news veracity label $\mathbf{y}$ is generated from both the graph $\mathcal{G}=\langle \mathbf{X}, \mathbf{A}\rangle$ and the environment $\mathbf{e}$. According to the dependence between variables $\mathbf{X}$, $\mathbf{A}$, $\mathbf{y}$, and $\mathbf{e}$, the joint probability $p(\mathbf{X}, \mathbf{A}, \mathbf{y}, \mathbf{e})$ can be given as: 
\begin{equation}
    p(\textbf{X}, \textbf{y}, \textbf{A}, \textbf{e}) = p(\textbf{e})p(\textbf{X})p(\textbf{A}|\textbf{X},\textbf{e})p(\textbf{y}|\textbf{X},\textbf{A},\textbf{e}),
\end{equation} 
where the generative models $p(\textbf{A}|\textbf{X},\textbf{e})$ and $p(\textbf{y}|\textbf{X},\textbf{A},\textbf{e})$ can be instantiated by flexible parametric distributions $p_{\theta}(\textbf{A}|\textbf{X},\textbf{e})$ and $p_{\theta}(\textbf{y}|\textbf{X},\textbf{A},\textbf{e})$ with parameter $\theta$. Most existing works aim to maximize the likelihood $\mathcal{P_{\theta}}(y|X, A)$, which is unsuitable for OOD prediction where $X$ and $A$ are causally affected by environment biases. 

We propose to filter the environment-biased information through data debiasing as follows. An environment variable $\mathbf{e}$ is defined as 1 or 0 for every training sample, indicating whether it is environment-independent ($\mathbf{e}=1$) or environment-biased ($\mathbf{e}=0$). Our training object is to optimize $\mathcal{P_{\theta}}(y|X_{e=1}, A_{e=1})$, to focus on environment-independent samples. 

To infer variable $\mathbf{e}$, posterior probability is utilized as follows: 
\begin{equation}
\label{Equation.posterior_prob}
\begin{split}
p_{\theta}(\textbf{e}|\textbf{A}, \textbf{X}, \textbf{y}) & = \frac{p_{\theta}(\textbf{X},\textbf{y},\textbf{A},\textbf{e})}{\sum_{\textbf{e'}\in\{0,1\}}p_{\theta}(\textbf{X},\textbf{y},\textbf{A},\textbf{e'})} \\
& = \frac{p(\textbf{e})p(\textbf{X})p_{\theta}(\textbf{A}|\textbf{X},\textbf{e})p_{\theta}(\textbf{y}|\textbf{X},\textbf{A},\textbf{e})}{\sum_{\mathbf{e'}\in\{0,1\}}p(\textbf{e'})p(\textbf{X})p_{\theta}(\textbf{A}|\textbf{X},\textbf{e'})p_{\theta}(\textbf{y}|\textbf{X},\textbf{A},\textbf{e'})},
\end{split}
\end{equation} 
where $p(\textbf{e})$ is the prior probability, which is set as a hyperparameter in the experiments; $p_{\theta}(\textbf{A}|\textbf{X},\textbf{e})$ is instantiated as the structure estimator (predicting edge connection $\textbf{A}$ based on features $\textbf{X}$ and environment $\textbf{e}$); and $p_{\theta}(\textbf{y}|\textbf{X},\textbf{A},\textbf{e})$ is instantiated as the classification module to predict the news veracity label, as detailed next. 

\section{Methodology}
\label{Section.methodology}
\begin{figure*}[tbh]
  \centering
  \includegraphics[width=0.75\linewidth]{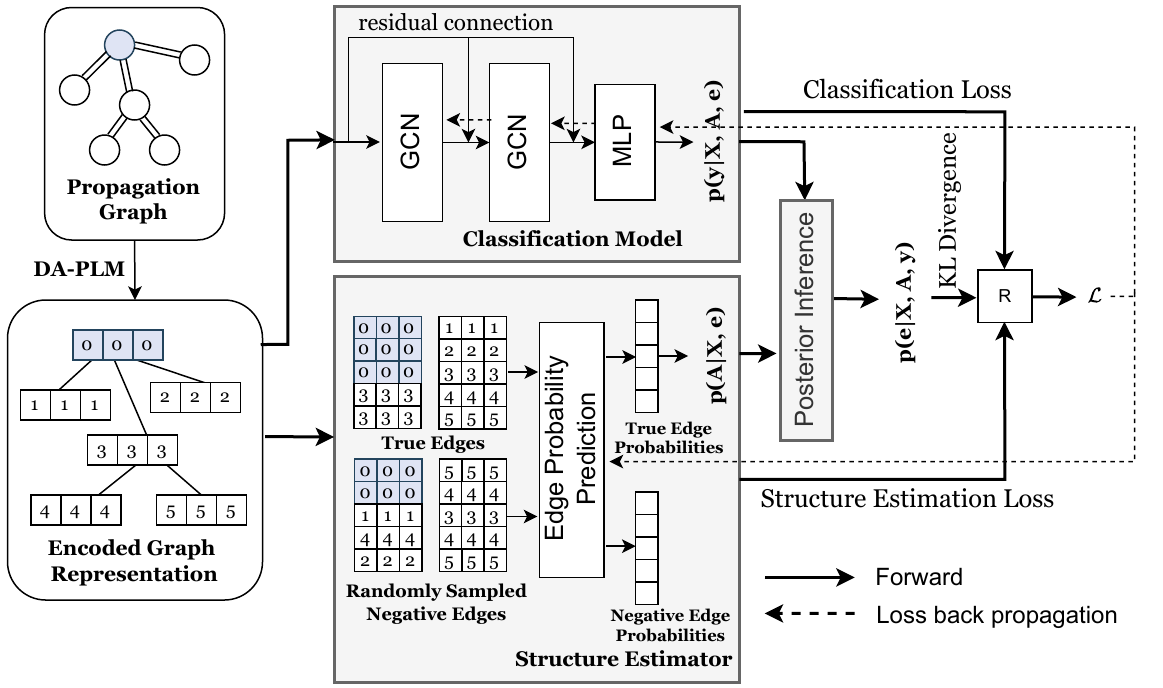}
  \caption{The structure of \method. R is the loss reweight module according to the inferred environment variable $\mathbf{e}$. }
  \Description{The model structure of~\method. }
  \label{Fig.model}
\end{figure*}

\textbf{Model overview.} 
The model structure (training phase) is shown in Fig.\ref{Fig.model}. Raw text data (news content and posts) are encoded using RoBERTa\cite{liu2019roberta}. The resulting propagation graph, with node feature embeddings, is input into a Classification Model and a Structure Estimator to generate the label prediction $p(\mathbf{y}|\mathbf{X}, \mathbf{A}, \mathbf{e})$ and connection likelihood $p(\mathbf{A}|\mathbf{X}, \mathbf{e})$. These outputs are used in posterior inference to estimate the environment variable $\mathbf{e}$, which weights the loss during training. The model is optimized via Expectation Maximization~\cite{moon1996expectation}. During testing, $\mathbf{e}$ is set to 1 for all samples, and $p(\mathbf{y}|\mathbf{X}, \mathbf{A}, \mathbf{e}=1)$ provides the final prediction.

\textbf{Classification model.} 
To instantiate the distribution $p_{\theta}(\mathbf{y}|\mathbf{X}, \mathbf{A}, \mathbf{e})$ we follow existing graph-based fake news detection models~\cite{bian2020rumor}, 
 combining a two-layer Graph Convolutional Network (GCN) and a multi-layer perception (MLP). Given a graph's node features $\mathbf{X} = \{\mathbf{x}_{1}, \mathbf{x}_{2}, \ldots, \mathbf{x}_{N}\}$ and its adjacency matrix $\mathbf{A}$, the propagation graph's embeddings are computed through GCNs with residual connections: 
\begin{equation}
\mathcal{Z}^{(l+1)} = \sigma\left(\mathbf{\tilde{D}}^{-1/2} \mathbf{\tilde{A}} \mathbf{\tilde{D}}^{-1/2} \mathcal{Z}^{(l)} \mathbf{W}^{(l)}\right) + \mathcal{Z}^{(l)},
\end{equation}
where $l=0$ or $1$; $\mathcal{Z}^{(0)}$ is the initial node features $\mathbf{X}$; $\mathbf{\tilde{A}}=\mathbf{A}+\mathbf{I}$ is the adjacent matrix of the graph with self-loops; $\mathbf{I}$ is the identity matrix; $\mathbf{\tilde{D}}$ is the degree matrix of $\mathbf{\tilde{A}}$; $\mathbf{W}^{(l)}$ is the learnable parameter matrix; and $\sigma$ is the activation function. After two layers of GCNs, the outputs $\mathcal{Z}^{(2)}$ is fed into the MLP to generate the prediction $\hat{y}$, where: 
\begin{equation}
\hat{y} = \text{softmax}(\text{MLP}(\mathcal{Z}^{(2)}))
\end{equation}

According to Equation~\ref{Equation.posterior_prob}, there are two probabilities for the classification model $p_{\theta}(\mathbf{y}|\mathbf{X},\mathbf{A},\mathbf{e}=1)$ and $p_{\theta}(\mathbf{y}|\mathbf{X},\mathbf{A},\mathbf{e}=0)$ representing the classification of the environment-independent and environment-biased samples, respectively. When the data is biased to environment, the correlation between the features and the veracity labels will be agnostic since the labels are more likely to be correlated to the environment biases (e.g.,  specific news events). Therefore, the above classification model only describes the probability $p_{\theta}(\mathbf{y}|\mathbf{X},\mathbf{A},\mathbf{e}=1)$. The instantiation of $e=0$ is introduced using Posterior Inference (see later).

\textbf{Structure estimator.}
Next, we instantiate $p_{\theta}(\mathbf{A}|\mathbf{X},\mathbf{e})$. Edge connections are estimated based on news contents $\mathbf{X}$ and environment variable $\mathbf{e}$. This is in line with reality  where posts are connected by interactions (comments/reposts), which can be inferred from the post contents (node features). For simplicity and following common practice~\cite{ma2019flexible}, we assume that the edges in the graph are conditionally independent. Then, the conditional probability of $\mathbf{A}$ can be factorized as $p_{\theta}(\mathbf{A}|\mathbf{X},\mathbf{e})=\prod_{i,j \in V}p_{\theta}(a_{ij}|\mathbf{X},\mathbf{e})$, where $p_{\theta}(a_{ij}|\mathbf{X},\mathbf{e})$ represents the probability of an edge existence between nodes $i$ and $j$ given node features $\mathbf{X}$. 

As with the classification model, when $\mathbf{e}=1$, the edge probability is inferred from the news content:
\begin{equation}
p_{\theta}(a_{ij}=1|x_{i}, x_{j}, \mathbf{e}=1) = \sigma([\mathbf{U}x_{i}, \mathbf{U}x_{j}]^{\top}\omega),
\end{equation}
where $\mathbf{U}$ and $\omega$ are learnable parameters, and $\sigma(\cdot)$ is the activation function. Since all edges already exist, to avoid model making trivial predictions as all edge probabilities being 1, we sample random edges from the propagation graph with the same number of positive edges for model training.

Same as with the classification model, the structure estimator only instantiates the scenarios for $e=1$, since for environment-biased samples, the propagation could express unstable or unknown patterns that should not be considered for classification. The instantiation of the structure estimator when $e=0$ is detailed next.

\textbf{Posterior inference.} 
\label{Section.Posteriror_inference} 
Using the instantiations of the classification model, the structure estimator $p_{\theta}(\mathbf{y}|\mathbf{X},\mathbf{A},\mathbf{e}=1)$ and $p_{\theta}(\mathbf{A}|\mathbf{X},\mathbf{e}=1)$ can be inferred. When $e=0$, to model handle the distribution of environment-biased samples, the classification model prediction is set to Gaussian distribution $\mathcal{N}(0,1)$, and the edge probability in the structure estimator $p_{\theta}(a_{ij}=1|x_{i}, x_{j}, \mathbf{e}=0)$ is set to $\mathcal{N}(0,1)$. Then, the probability of a sample being environment-independent can be inferred from Equation~\ref{Equation.posterior_prob} with the preset prior $p(e)$, which is provided as a hyperparameter.

\textbf{Training objectives.} 
Based on the instantiation of the classification model and the structure estimator, the environment variable can be inferred according to Equation~\ref{Equation.posterior_prob} as shown by the posterior inference part of Fig.~\ref{Fig.model}. 

In the training process, we are given the graph node features $\textbf{X}$, edge connections $\textbf{A}$ and labels $\textbf{y}$. 
Parameter $\theta$ is used to perform news classification, structure estimation and environment variable inference (data debiasing). 
The model is trained by optimizing the Evidence Lower BOund (ELBO) of observed data tuple $(\textbf{A},\textbf{X},\textbf{y})$ based on Equation~\ref{ELBO}.

\begin{equation}
\label{ELBO}
\begin{split}
\log {p_{\theta}(A,y|X)}  \ge \log_{}{p_{\theta}(A, y|X)} - D_{KL}[p_{\theta}(\mathbf{e}|A, X, y)||p(e)] \\
=E_{p_{\theta}(e|A,X,y)}[\log_{}{p_{\theta}(A|X,e)p_{\theta}(y|X,A,e)}] \\
- D_{KL}[p_{\theta}(\mathbf{e}|A, X, y)||p(e)] = \mathcal{L}_{ELBO}
\end{split}
\end{equation}

The final learning objective is the sum of three terms: 
(1) the classification loss $\mathcal{L}_{cl}$ shown in Equation~\ref{Equation.L_cl};
(2) the structure regularization loss $\mathcal{L}_{reg}$ shown in Equation~\ref{Equation.L_reg}; and 
(3) the KL divergence loss $\mathcal{L}_{KL}$ between the estimated environment variable and the prior shown in Equation~\ref{Equation.L_kl}. 

\begin{equation}
\label{Equation.L_cl}
\begin{split}
\mathcal{L}_{cl} = -\frac{1}{N^{tr}}\sum_{i=1}^{N_{tr}}E_{p_{\theta}(e_{i}|A_{i},X_{i},y_{i})}& [e_{i}\log{p_{\theta}(y_{i}|X_{i},A_{i},e_{i}=1)}+ \\ 
& (1-e_{i})\log{p_{\theta}(y_{i}|X_{i},A_{i},e_i=0)}] 
\end{split}
\end{equation}

\begin{equation}
\label{Equation.L_reg}
\begin{split}
\mathcal{L}_{reg} =  - \frac{1}{N^{tr}}\sum_{i=1}^{N_{tr}}E_{p_{\theta}(e_{i}|A_{i},X_{i},y_{i})} [e_{i}\log{p_{\theta}(A_{i}|X_{i},e_{i}=1)}+ \\  (1-e_i)\log{p_{\theta}(A_{i}|X_{i},e_{i}=0)}] \\
\end{split}
\end{equation}

\begin{equation}
\label{Equation.L_kl}
\mathcal{L}_{KL} = - \frac{1}{N^{tr}}\sum_{i=1}^{N_{tr}}D_{KL}[p_{\theta}(\mathbf{e_{i}}|A_i, X_i, y_i)||p(e)] 
\end{equation}


\section{Experiment}

To evaluate our model performance on unseen fake news detection, an OOD fake news detection benchmark is used following the approach given in~\cite{ran-aaai-2023-unsupervised}. The datasets used to train the models and the datasets used to test the models are from non-overlapping news domains. At training, no knowledge of test data is leaked, including news content, propagation graphs, and veracity labels, except for the UCD-RD~\cite{ran-aaai-2023-unsupervised} model, which will use features of the test data to adjust model parameters through contrastive learning. 


\textbf{Datasets.} 
Four public datasets collected from Twitter (now called X) and Weibo (a Chinese social media platform like Twitter) are used. They are \texttt{Twitter} \cite{ma2017detect}, \texttt{Weibo}~\cite{ma2018rumor}, \texttt{Twitter-COVID19}~\cite{lin-naacl-2022-aclr} and~\texttt{Weibo-COVID19} \cite{lin-naacl-2022-aclr}. 
\texttt{Twitter} and \texttt{Weibo} comprise news/posts from general domains (named open-domain), and are treated as the training set.  \texttt{Twitter-COVID19} and \texttt{Weibo-COVID19}  only contain news/posts related to COVID-19. They represent data from an emerging/unseen topic. The statistics of the datasets are shown in Table~\ref{Tab.data-stats}.

\begin{table}[h!]
\centering
\caption{Dataset Statistics (`\#' means ``number of'', `Avg.' means  average).}
\label{tab:statistics}
\begin{tabular}{@{}lcccc@{}}
\toprule
\textbf{Statistics} & \textbf{Twitter} & \textbf{T-COVID} & \textbf{Weibo} & \textbf{W-COVID} \\
\midrule
\# events           & 1,154    & 400    & 4,649   & 399    \\
\# tree nodes       & 60,409   & 406,185  & 1,956,449 & 26,687\\
\# true news       & 579     & 148    & 2,336   & 146 \\
\# fake news          & 575     & 252    & 2313   & 253    \\
Avg. lifetime  & 389 Hrs & 2,497 Hrs & 1,007 Hrs & 248Hrs \\
Avg. depth/tree        & 11.67   & 143.03   & 49.85  & 4.31\\
Avg. \# posts  & 52      & 1,015     & 420    & 67 \\
Domain                 & Open    & COVID-19 & Open   & COVID-19 \\
Language               & English & English & Chinese & Chinese \\
\bottomrule
\end{tabular}
\label{Tab.data-stats}
\end{table}

Detection is performed in both cross-domain and cross-language settings to evaluate the models' generalization capability. When testing performance on \texttt{Weibo-COVID19} in Chinese, the models are trained on \texttt{Twitter} in English, and similarly for  \texttt{Twitter-COVID19}.

\textbf{Baselines.} To evaluate the model, 
baselines include sequence-based models \textbf{LSTM}~\cite{ma2016detecting}, \textbf{RvNN}~\cite{ma2018rumor}, Transformer-based models \textbf{PLAN}~\cite{khoo2020interpretable}, \textbf{RoBERTa}~\cite{liu2019roberta}, graph-based models \textbf{BiGCN}~\cite{bian2020rumor}, \textbf{GACL}~\cite{sun2022rumor}, \textbf{SEAGEN}~\cite{gong2023fake} and domain-adaptive model \textbf{UCD-RD}~\cite{ran-aaai-2023-unsupervised} are experimented.

\begin{table}[htb!]
    \small
    \caption{Zero-Shot Fake News Detection  on \texttt{Twitter-COVID19} and \texttt{Weibo-COVID19} (Acc: Accuracy; F-F1: F1 score on fake news detection; T-F1: F1 score on true news detection).}
    \centering
    \setlength{\tabcolsep}{1.78mm}{
    \begin{tabular}{l | c c c| c c c }

        \toprule
        Source & \multicolumn{3}{c|}{\texttt{Twitter}}& \multicolumn{3}{c}{\texttt{Weibo}} \\ \midrule
        Target & \multicolumn{3}{c|}{\texttt{Weibo-COVID19}} & \multicolumn{3}{c}{\texttt{Twitter-COVID19}}\\ \midrule
        Method & Acc &  T-F1  & F-F1 & Acc & T-F1  & F-F1 \\\midrule

        LSTM   & 0.463 & 0.329 & 0.498 & 0.510 & 0.243 & 0.533 \\
        RvNN   & 0.514 & 0.426 & 0.538 & 0.540 & 0.247 & 0.534 \\
        PLAN   & 0.532 & 0.414 & 0.578 & 0.573 & 0.298 & 0.549 \\
        RoBERTa& 0.623 & 0.459 & \underline{0.711} & 0.603 & \textbf{0.585} & 0.619 \\ \midrule
        BiGCN  & 0.569 & 0.429 & 0.586 & 0.616 & 0.252 & 0.577 \\
        SEAGEN & 0.555 & 0.406 & 0.583 & 0.578 & 0.320 & 0.650 \\
        GACL   & 0.601 & 0.410 & 0.616 & \underline{0.621} & 0.345 & \underline{0.666} \\ 
        UCD-RD & \underline{0.631} & \underline{0.510} & 0.621 & 0.591 & 0.371 & 0.583 \\ 
         \method & \textbf{0.754} & \textbf{0.620} & \textbf{0.819} & \textbf{0.693} & \underline{0.513} & \textbf{0.775} \\
        \hline
        $\uparrow$ (\%) & +19.49 & +21.57 & +15.19 & +11.59 & -12.31 & +16.37 \\ \bottomrule

    \end{tabular}}
    \label{Tab.res1}
\end{table}

\textbf{Implementation.} 
All baselines and the \method~ model are implemented using PyTorch\footnote{https://pytorch.org/} and trained with an NVIDIA A100 80 GB GPU. The baseline models use default hyperparameter settings from their papers. Hyperparameter $p(e)$ of \method~ indicates the proportion of environment-independent samples. This is set to 0.7 and 0.6 in the source-Twitter and source-Weibo experiments, respectively. Our source code will be released upon paper publication.

\textbf{Results.} 
The experiment results are shown in Table~\ref{Tab.res1}. As seen, the sequence-based methods have the worst performance in both accuracy and F1 score due to their limited feature extraction capability. The graph-based models generally perform better than the sequence-based ones, highlighting the effectiveness of propagation graphs, with the exception of SEAGEN where the performance drops compared to the reported results in its original paper. This

\noindent may be due to the temporal features that it uses also suffering the distribution shift, i.e., the news originating from different domains has different temporal features. Leveraging data augmentation, contrastive learning and testing features,  GACL and UCD-RD's performances are among the best in the baselines. Our \method's superior performance demonstrates the effectiveness of causal debiasing, even though we only use a simple two-layer GCN encoder to encode the propagation graphs.

\begin{figure}[htb]
  \centering
  \begin{subfigure}[b]{0.49\linewidth}
    \centering
    \includegraphics[width=\linewidth]{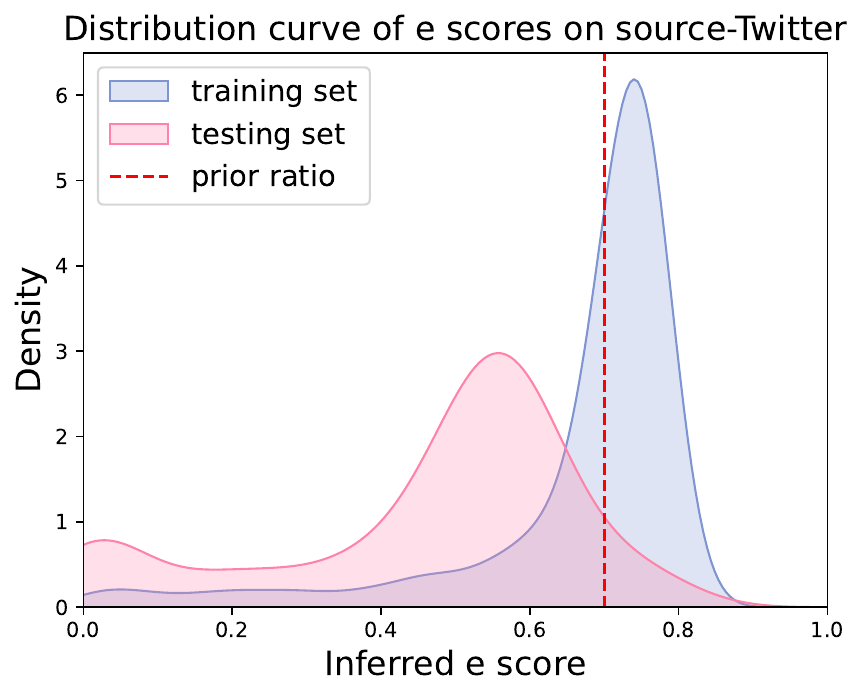}

    \label{Fig.twitter_distribution}
  \end{subfigure}
  \hfill 
  \begin{subfigure}[b]{0.49\linewidth}
    \centering
    \includegraphics[width=\linewidth]{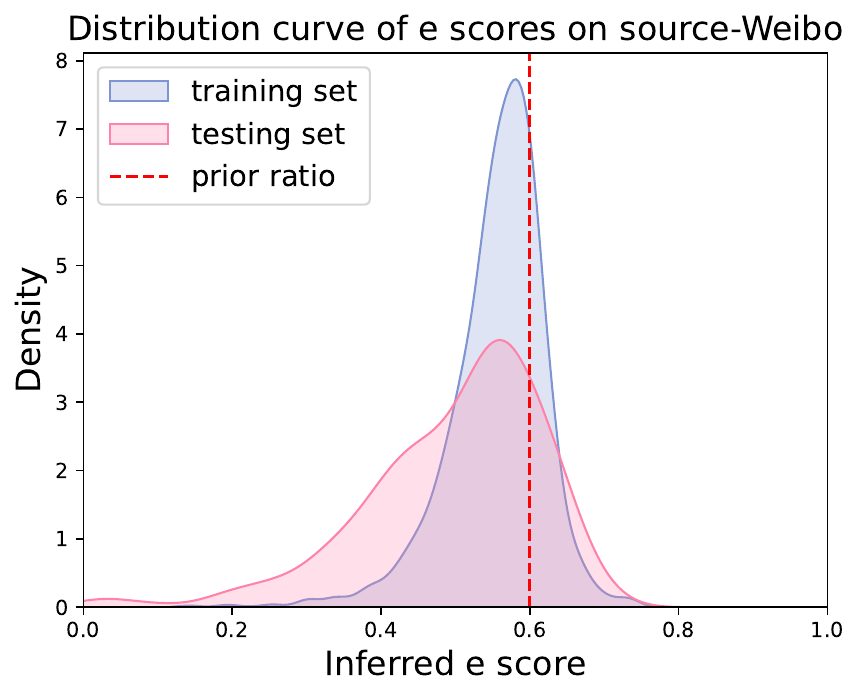}

    \label{Fig.weibo_distribution}
  \end{subfigure}

  \caption{Distribution of inferred environment variable (left: source Twitter dataset, right: source Weibo dataset).}
  \Description{The distribution curve of the inferred environment variable for both the Twitter and Weibo datasets.}
  \label{Fig.distribution_curve_e}
\end{figure}

\textbf{Case study.} 
The environment inference results are shown in Fig.~\ref{Fig.distribution_curve_e}. The distribution of inferred environment variable $e$ is plotted. As can be seen, the majority of training samples are assigned weights around the prior ratio. We can treat the samples with weights far away from the prior ratio as environment-biased samples. From the analysis of the model function, these samples are either difficult to classify or their propagation is hard to estimate. We also find that debiasing using the test samples is actually drawing the latent distributions of the training and testing samples closer together in an unsupervised way.

\begin{figure}[tbh]
  \centering
  \includegraphics[width=0.75\linewidth]{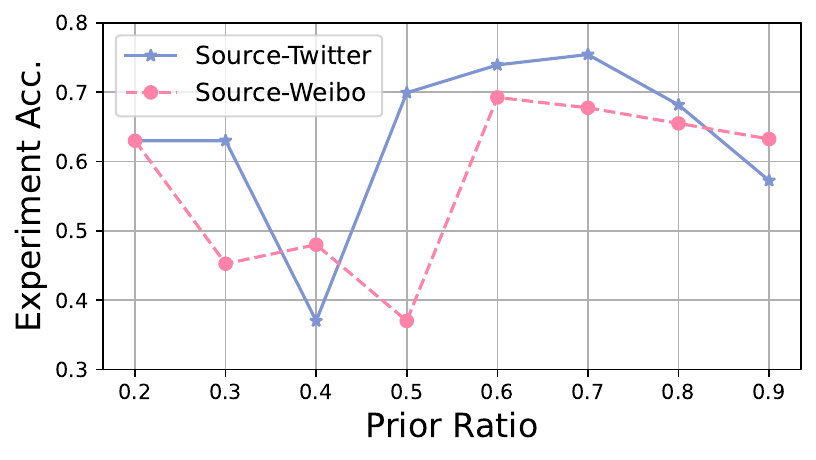}
  \caption{Parameter $p(e)$ sensitivity.}
  \Description{Parameter $p(e)$ sensitivity. }
  \label{Fig.parameter_sensitivity}
\end{figure}

\textbf{Parameter study.} 
The hyperparameters of the prior distribution $p(e)$ are selected through grid search, as shown in Fig.~\ref{Fig.parameter_sensitivity}. The results highlight the importance of a reasonable prior: a value set too high assumes all training data is environment-independent, reducing the debiasing effect, while a value set too low treats all training data as biased, leading to model underfitting.

\section{Related Work} 

The detection of fake news has been extensively studied with existing solutions typically relying on news content analysis, propagation structures, or user credibility assessment. However, such solutions often struggle when faced with distributional shifts between training and testing data, leading to performance degradation. To address this issue, researchers have explored cross-domain fake news detection, which focuses on training a model in one domain (the \emph{source domain}) and applying it to another (the \emph{target domain}). Broadly, these solutions can be classified into \emph{sample-level} and \emph{feature-level} approaches.  

Sample-level approaches aim to identify training samples that exhibit domain-invariant characteristics, assigning them greater importance during model training~\cite{silva2021embracing,yue-cikm-2022contrastive}. Some studies~\cite{yue-cikm-2022contrastive,ran-aaai-2023-unsupervised} enhance target domain data by employing clustering algorithms to generate augmented samples, which are then incorporated into training alongside source domain data. This strategy strengthens model robustness when handling unseen domains. Feature-level approaches, on the other hand, focus on identifying and emphasizing domain-independent attributes. For example, reinforcement learning has been applied to select features that remain stable across domains~\cite{mosallanezhad2022domain}. Inspired by domain-adaptive learning techniques~\cite{ganin-icml-2015-dann-unsupervised}, some works~\cite{min2022divide,li-acmmm-2023improving} train a domain discriminator adversarially to encourage the model to generate news embeddings that obscure domain-specific characteristics, thereby improving generalization.  

Our approach takes a step further by leveraging causal analysis on the propagation structure, capturing more informative patterns that contribute to cross-domain fake news detection.

\section{Conclusions and Limitations}
We demonstrated that \method~achieves state-of-the-art performance in detecting unseen fake news by addressing domain-specific biases in training data through causal analysis and reweighting strategies. Besides, the reweighing strategy is only applied during the training process, leaving the trained graph encoders and classifier for the testing (veracity inference) process, to improve the scalability of real-application. 

Our work has certain limitations, such as the need to pre-define the prior ratio, which could be enhanced by dynamically estimating a pseudo-environment variable. Additionally, the dependency between textual content and the environment could be more effectively modeled. Leveraging the capabilities of LLMs could address this challenge, as they have been utilized in fake news detection both as supportive agents~\cite{hu2024bad} and as advanced news content processors~\cite{ma2024fake}. Future research could also explore scalable, real-time fake news detection, ideally in collaboration with social media platforms like Twitter/X.

\begin{acks}
This study is supported by Melbourne Research Scholarship and CSIRO Data61 Top-up Scholarship. 
\end{acks}

\bibliographystyle{ACM-Reference-Format}
\bibliography{ref}

\appendix

\end{document}